\documentclass{aa} % ``standard'' version
\usepackage{graphicx}
\usepackage{epsfig}
\usepackage{natbib}
\bibpunct{(}{)}{;}{a}{}{,}
\begin{document}

%   \thesaurus{06     % A&A Section 6: Form. struct. and evolut. of stars
%              (03.13.2;  % Methods: data analysis,
%               02.12.2)} % Line: identification,
%
%   \thesaurus{15     % astronomical instrumentation, methods and techniques
%              (15.13.2;  % Methods: data analysis,
%               02.12.2)} % Line: identification,
%

\title{
%Identification and modelling of early mission SPI background
%lines -- DRAFT -- !!!
%First (???) view of the all-sky distribution of positronium annihilation
%continuum emission with SPI/INTEGRAL
The sky distribution of positronium annihilation continuum emission
measured with SPI/INTEGRAL\thanks{Based on observations with INTEGRAL,
an ESA project with instruments and science data centre funded by ESA
member states (especially the PI countries: Denmark, France, Germany,
Italy, Switzerland, Spain), Czech Republic and Poland, and with the
participation of Russia and the USA.}}

%   \subtitle{I. Overviewing the $\kappa$-mechanism}

\author{G. Weidenspointner\inst{1}\thanks{External ESA Fellow}
        \and
	C.R. Shrader\inst{2}
        \and
        J. Kn{\"o}dlseder\inst{1}
        \and
        P. Jean\inst{1}
        \and
        V. Lonjou\inst{1}
        \and
	N. Guessoum\inst{3}
	\and
	R. Diehl\inst{4}
	\and
	W. Gillard\inst{1}
	\and
        M.J. Harris\inst{1}
        \and
	G.K. Skinner\inst{1}
        \and
        P. von Ballmoos\inst{1}
	\and
	G. Vedrenne\inst{1}
	\and
	J.-P. Roques\inst{1}
	\and
	S. Schanne\inst{5}
	\and
	P. Sizun\inst{5}
	\and
	B.J. Teegarden\inst{2}
        \and
	V. Sch\"onfelder\inst{4}
        \and
	C. Winkler\inst{6}
       }

\offprints{G. Weidenspointner, e-mail Georg.Weidenspointner@cesr.fr}

\institute{Centre d'Etude Spatiale des Rayonnements, 9 Avenue Colonel
           Roche, 31028 Toulouse Cedex 4, France
           \and
           NASA Goddard Space Flight Center, LHEA, Code 661, Greenbelt,
           MD 20771, USA
           \and
           American University of Sharjah, College of Arts \& Science,
Physics Department, PO Box 26666, Sharjah, UAE
           \and
           Max-Planck-Institut f\"ur extraterrestrische Physik,
Postfach 1603, 85740 Garching, Germany
           \and
           DSM/DAPNIA/SAp, CEA Saclay, 91191 Gif-sur-Yvette, France
           \and
           ESA/ESTEC, Science Operations and Data Systems Division
(SCI-SD), 2201 AZ Noordwijk, The Netherlands
%           \and
%           Universities Space Research Association, 7501 Forbes
%           Blvd. \#206, Seabrook, MD 20706, USA
%           \and
%           C.S.N.S.M., IN2P3-CNRS and Universite Paris-Sud, F-91405
%           Orsay, France
%           \and
%           DSM/DAPNIA/SAp, Centre d'Etudes Nucleaires de Saclay,
%           91191-GIF-sur-Yvette CEDEX, France
%           \and
%           Max-Planck-Institut f{\"u}r extraterrestrische Physik,
%           Giessenbachstrasse, 85740 Garching, Germany 
%           \and
%           University of New Mexico, Albuquerque, NM 87131, USA
%           \and
%           Department of Physics and Astronomy, University of
%           Southampton, Southampton, SO17 1BJ, UK
%           \and
%           Institut de Physique Nucleaire, Universite catholique de
%           Louvain, B-1348 Louvain-la-Neuve, Belgium
          }

\date{Received ; accepted }

% because title is too long

   \titlerunning{The sky distribution of positronium continuum
emission}
   \authorrunning{G. Weidenspointner et al.}

%%%   abstract   %%%

\abstract{
We present a measurement of the sky distribution of positronium
(Ps) annihilation continuum emission obtained with the SPI
spectrometer on board ESA's INTEGRAL observatory. The only sky region
from which significant Ps continuum emission is detected is
the Galactic bulge.
%Emission from the Galactic disk is at best marginally detected
%(depending on the Galactic disk model) and can be expected to be
%mainly Galactic diffuse continuum emission. 
The Ps continuum emission is 
%spherically symmetric 
%{\bf spherically/azimuthally ???} 
circularly symmetric about
%and centred on 
the Galactic centre, with an extension of about $8^\circ$ FWHM. Within
measurement uncertainties, the sky distribution of the Ps continuum
emission is consistent with that found by us for the 511~keV
electron-positron annihilation line using SPI. Assuming that 511~keV
line and Ps continuum emission follow the same spatial distribution,
we derive a Ps fraction of $0.92\pm0.09$. These results strengthen our
conclusions regarding the origin of positrons in our Galaxy based on
observations of the 511~keV line.  In particular, they suggest that
the main source of Galactic positrons is associated with an old
stellar population, such as Type~Ia supernovae, classical novae, or
low-mass X-ray binaries. Light dark matter is a possible alternative
source of positrons.
}

\maketitle

%
%________________________________________________________________

\section{\label{introduction} Introduction}

The annihilation of positrons with electrons gives rise to two
characteristic emissions at gamma-ray energies: the hallmark line at
511~keV, and the unique three-photon positronium (Ps) continuum
emission \citep[cf.][]{Guessoum2005}. 
%
%The triangular Ps continuum peaks at 511~keV and is produced by
%annihilation of positrons and electrons that first form ortho-Ps with
%their spins being parallel. If the spins of positron and electron are
%anti-parallel after forming para-Ps, their annihilation gives rise to
%two 511~keV photons, as is the case for direct annihilation (i.e.\
%without prior formation of Ps).
%
Direct annihilation of positrons with electrons, and their
annihilation via the formation of para-Ps (with the spins of electron
and positron being anti-parallel), result in the emission of two
511~keV photons. Annihilation via the formation of ortho-Ps (with the
spins of electron and positron being parallel) produces three photons
and gives rise to the Ps continuum emission, which is roughly
saw-tooth shaped with a peak at the maximum energy of 511~keV
\citep{Ore_Powell49}. 

%Cosmic positron annihilation radiation was first detected in
%balloon observations from the Galactic centre (GC) direction in the 1970s
Cosmic positron annihilation radiation was first detected from the
Galactic centre (GC) direction in balloon observations during the 1970s
and has been the focus of intense scrutiny by a large number of
balloon and satellite borne experiments ever since
%\citep[see reviews by, e.g.,][]{lingenfelter_ramaty89, harris97}.
\citep[see e.g.\ the reviews by][]{Tueller92,Harris97}.
Despite tremendous observational and theoretical efforts, the origin
of the positrons is still poorly understood.
A large variety of positron sources and production mechanisms have
been proposed over the years 
%\citep[see e.g.\ discussions in][]{chan_lingenfelter93,
%ramaty_lingenfelter94}
\citep[e.g.][]{Chan_Lingenfelter93, Dermer_Murphy01}. Among the more
promising source candidates are radioactive nucleosynthesis products from
supernovae. More recently, hypernovae/GRBs
\citep{Casse04, Parizot05} and light 
%or decay 
dark matter 
%or relic particles \citep[e.g.][]{boehm04, hooper_wang04} have been
\citep[e.g.][]{Boehm04} have been
proposed as possible candidates. Compact objects comprise another
potential candidate source of positrons.

Investigations of the 
%morphology 
sky distribution of the annihilation radiation promise to provide
clues to the identification of the source(s) of positrons in our
Galaxy, despite the fact that positrons may travel from their birth
places before annihilating. First maps of the annihilation radiation,
limited to the inner regions of our Galaxy, were obtained using the
OSSE instrument on board the Compton Gamma-Ray Observatory
\citep{Johnson93} in the
511~keV line and in Ps continuum emission \citep[e.g.][]{Purcell97,
Chen97, Milne00, Milne01a, Milne01b, Milne02}. Furthermore, the OSSE
instrument allowed
\citet{Kinzer99, Kinzer01} to study the one-dimensional distribution
in longitude and in latitude of diffuse emission, including annihilation
radiation, from the inner Galaxy.
%
%The imaging spectrometer SPI
%%\citep{Vedrenne03} 
%on board ESA's INTEGRAL observatory allowed different groups 
%to perform improved mapping of the 511~keV line
%\citep[e.g.][]{Weidenspointner04, Knoedlseder05}. 
%Improved mapping of the 511~keV line has become feasible with the
%commissioning of the imaging spectrometer SPI on board ESA's INTEGRAL
%observatory \citep{Jean03a, Weidenspointner04, Knoedlseder05}. 
%A first search for large-scale 511~keV line emission from the Galactic
%plane using SPI was presented by \citet{Teegarden05}.
With the commissioning of the imaging spectrometer SPI on board ESA's
INTEGRAL observatory, high spectral resolution mapping with improved
angular resolution has become feasible \citep{Jean03a,
Weidenspointner04, Knoedlseder05}.
The 511~keV line emission is found to be dominated by the Galactic
bulge and/or halo; emission from the Galactic disk is much fainter,
implying that positron annihilation is concentrated in the central
regions of our Galaxy \citep{Knoedlseder05}.

A first SPI measurement of the flux in Ps continuum
emission, and of the Ps fraction $f_{Ps}$ \citep[the fraction of
positrons that annihilate through Ps
formation;][]{Brown_Leventhal87}, has been presented by
\citet{Churazov05}. The value of $f_{Ps}$, as well as the
detailed shape of the 511~keV annihilation line, depend on the
physical properties of the annihilation media; therefore detailed
spectroscopy of the positron annihilation can provide unique
information on the annihilation media and processes
\citep{Guessoum2005, Jean06}.

In this publication, we present results concerning the Galactic
distribution of Ps continuum emission using observations of most of
the celestial sphere with the spectrometer SPI/INTEGRAL. The analysis
of the SPI observations is presented in Sec.~\ref{data_analysis}. Our
mapping, model fitting, and spectral results are given in
Sec.~\ref{results}. A summary and our conclusions can be found in
Sec.~\ref{summary}.

%\section{Bla}
%
%The Spectrometer for INTEGRAL (SPI) is one of the two main instruments
%on board ESA's INTEGRAL observatory launched from Baikonour,
%Kazakhstan, on Oct.~17, 2002. The INTEGRAL mission was placed into a
%highly elliptical orbit with a perigee of 9000~km. Consequently,
%INTEGRAL does not benefit from geomagnetic shielding and is fully
%exposed to all cosmic rays. Interactions of these cosmic rays within
%the instrument and spacecraft materials are the dominant source of
%instrumental background for SPI. In particular, delayed decays of
%radio-isotopes and prompt de-excitations of excited nuclei produced in
%nuclear interactions give rise to a plethora of instrumental lines
%which are the focus of this work. The general characteristics of the
%SPI instrumental background and its temporal and orbital variation are
%described by \citet{jean_spi_bgd}.
%
%A detailed understanding of the instrumental lines is valuable for both
%the operation of the instrument as well as for scientific
%analyses. Accurate line identifications are a prerequisite for the
%absolute energy calibration of the detectors and the monitoring of their
%radiation damage. Many scientific analyses, in particular studies of
%diffuse gamma-ray emission from the Galaxy, necessitate modelling both
%the amplitude and shape of the instrumental background in specific
%energy regions. Typically, this involves modelling of instrumental
%lines.

%
%________________________________________________________________

\section{\label{data_analysis} Instrument description and data analysis}

The SPI imaging spectrometer consists of an array of 19 actively
cooled high resolution Ge detectors, 
% with a total volume of 3396~cm$^3$. The detectors 
which cover an energy range of 20--8000~keV with an energy resolution
of about 2.1~keV 
%full width at half maximum 
FHWM at 511~keV. SPI
employs an active anti-coincidence shield made of bismuth germanate
(BGO), which also acts as a collimator. In addition to its
spectroscopic capabilities, SPI can image the sky with moderate spatial
resolution of about 3$^\circ$ FWHM using a tungsten coded aperture
mask. The fully coded field-of-view of the instrument is about
$16^\circ$. 
% located 1.71~m above the detector array.  
A detailed description of the instrument was given by
\citet{Vedrenne03}.

%Vedrenne et al.\ \cite{vedrenne_spi_instr},
%Attie et al.\ \cite{attie_spi_cal}, and Roques et 
%al.\ \cite{roques_spi_perf}.

The analysis presented here is based on the same data set that we used
to study the sky distribution of the 511~keV electron-positron
annihilation line \citep[for details see][]{Knoedlseder05}. The data
comprise all observations included in the December 10, 2004 public
INTEGRAL data release (i.e.\ three-day orbital revolutions 16--76,
79--80, 89--122) supplemented by instrument team observations of the
Vela region during revolutions 81--88. The observations were taken
during the epoch December 9, 2002 through September 12, 2003. In
total, the data set consists of 6821 pointings\footnote{Each INTEGRAL
observation is split into numerous so-called pointings, often arranged
on a square grid of 25 positions on the sky centred on the main
target of the observation, which each pointing lasting about 2000~s
\citep{Courvoisier03}. This so-called dithering strategy is necessary
to minimize imaging artefacts for SPI by providing a sufficient number of
measurements (sky signal and instrumental background) for each of the
19 detectors to constrain sufficiently the fluxes from each pixel in
the field-of-view.} with a
combined 
%observation time of $15.3 \times 10^6$~s. 
live time of $13.5 \times 10^6$~s. The resulting
exposure to the sky \citep[see Fig.~1 in][]{Knoedlseder05} is
relatively uniform for Galactic longitudes $|l| \le 50^\circ$ and
Galactic latitudes $|b| \le 15^\circ$, with a maximum at the GC
region. Unexposed regions are limited to the south Galactic pole
region and a few areas at intermediate Galactic latitudes.

The analysis of the Ps continuum emission, and of the underlying
Galactic power-law continuum (hereafter: Galactic continuum)
emission\footnote{This emission is partly due to cosmic-ray
interactions in the interstellar medium, partly due to point
sources.}, is rendered challenging by a signal-to-background ratio
which is only of the order of a few tenths of one per cent (compared
to a few per cent for the 511~keV annihilation line). Background
modelling therefore is crucial.
%
%It is done by fitting time series of backgrounds and of
%source exposures to time series of data. 
%We followed the approach described in detail in
%\citet{Knoedlseder05}. 
Our approach consists of fitting time series of detector count rates
by a linear combination of so-called background templates (time series
of background components, explained below). The detector signal
arising from a celestial source is also variable in time, even if the
source is intrinsically stable, because SPI's exposure to it varies in
time as the instrument performs a series of observations. 
%{\bf When
%analyzing extended emission, ideally these observations include sky
%regions devoid of the emission of interest.} 
Source
components can be distinguished from the dominant background
components by taking advantage of their differing time
series.
%\footnote{\bf For extended emission subtending at least a
%signifcant fraction of the field-of-view it is not possible to
%extract its flux or its spectrum from a single observation. The small
%number of pointings (usually 25 per observation) does not provide
%enough information to separate signal from background. It is therefore
%necessary to combine many observations.}.

Many processes contribute to
the instrumental background at all energies
% around and below 511~keV
\citep{Jean03b, Weidenspointner03}. The statistical
significance of the available data is still a limiting factor in the
analysis of sky signals.  We found that the background components used
by \citet{Knoedlseder05} to model the instrumental 511~keV line
provide a viable compromise between accuracy and simplicity. The model
consists of three components: a constant, the rate of saturating
($>8$~MeV) events in the Ge detectors, and a template representing the
exponential build-up of long-lived isotopes (a lifetime of 352~days,
corresponding to $^{65}$Zn, was chosen\footnote{Since our data extend
only over about 1~year, we cannot yet distinguish different half-lives
longer than about 6 months. The exact value of the lifetime is
therefore not crucial,
%it is only the presence of a (any) long-lived component that
%is important.
but it is important to allow for the significant long-term build-up
effect.}). The role of the first two templates is to account for
prompt and short-lived background components, the third template
accounts for long-lived background components.
% \citep[for details, see][]{Knoedlseder05}. 
This three-component background model was applied either separately to
three energy intervals, namely 410--430, 447--465, and 490--500~keV
(hereafter: Ps analysis intervals), or to the summation of the events
in these three intervals into a single data set (hereafter: summed Ps
analysis intervals). The intervals were chosen to be free of strong
background features (which require specific modelling using other,
intermediate time scales).
%The analysis of the sky distribution of
%positronium emission was therefore performed in three energy
%intervals: 
We do not include the fourth background component used in the
511~keV line analysis, which represented the continuum underlying the
instrumental 511~keV line.
% \citep{Knoedlseder05}. 
Thus we are measuring the total emission, Ps continuum plus Galactic
continuum.
%

%{\bf ??? state somewhere else ???} A major difficulty for the analysis
%of Ps continuum emission arises from the fact that the spatial
%distribution of the underyling Galactic continuum emission is still
%poorly constrained at soft gamma-ray energies.

As in studies of the sky distribution of the 511~keV line emission, we
found that by fitting the normalizations of each of the three
background model components for each detector remaining systematic
uncertainties introduce a bias.  However, if we fit the second
background component not only for each SPI detector, but additionally
for each orbital revolution
\citep[as in model {\tt ORBIT-DETE} in][]{Knoedlseder05} we reduce
systematic trends to an acceptable level. This method is similar to
those applied by us in earlier analyses of the 511~keV line
\citep[e.g.][]{Jean03a, Knoedlseder03, Weidenspointner04}. 

To obtain model independent maps of extended emission from our Galaxy
we employed an implementation of the Richardson-Lucy algorithm that is
described in detail in \citet{Knoedlseder05}. Briefly, in this
approach the intensity attributed to each sky pixel is calculated
iteratively using the relation
\begin{equation}
\label{RL_iteration}
f_j^{k+1} = f_j^k + \lambda^k \omega_j f_j^k \, 
\left( \frac{\sum_{i=1}^N \left(\frac{n_i}{e_i^k}-1\right) \,
R_{ij}}{\sum_{i=1}^N R_{ij}}\right)
\end{equation}
where $k$ is the iteration number, $f_j^k$ is the intensity from sky
pixel $j$ in iteration $k$, $R_{ij}$ is the instrumental response
matrix (linking data space bin $i$ to sky pixel $j$), $n_i$ is the
number of counts measured in data space bin $i$, $e_i^k = \sum_{j=1}^M
R_{ij} j_j^k + b_i$ is the predicted number of counts in data space
bin $i$ after iteration $k$ (with $b_i$ being the predicted number of
instrumental background counts for data space bin $i$), $N$ and $M$
are the dimensions of the data and image spaces, respectively, and
$\lambda^k$ is an acceleration factor. To avoid noise artefacts in
weakly exposed regions of the sky, the image increments are weighted
with a quantity $\omega_j = (\sum_{i=1}^N R_{ij})^2$ that is related
to the sensitivity of the instrument.

When fitting models of intensity distributions on the sky to the
data, we employed the maximum likelihood multi-component model fitting
algorithm described in detail in \citet{Knoedlseder05}. Briefly,
assuming Poisson noise for the measured number $n_i$ of events in each
of the $N$ data space bins, the algorithm maximizes the log likelihood
function
\begin{equation}
\label{log_likelihood}
\ln L = \sum_{i=1}^N n_i \ln e_i - e_i - \ln n_i!
\end{equation}
where $e_i = \sum_k \alpha_k s_i^k + b_i(\mbox{\boldmath$\beta$})$ is
the predicted number of (source plus background) counts in data space
bin $i$, $s_i^k = \sum_{j=1}^M f_j^k R_{ij}$ is the sky intensity
model $k$ folded into data space bin $i$ (with $f_j^k$ being the intensity
in sky pixel $j$ due to sky model $k$ and $R_{ij}$ being the
instrumental response matrix), $b_i(\mbox{\boldmath$\beta$})$ is the
background model, and $\alpha_k$ and $\mbox{\boldmath$\beta$}$ are
scaling factors for the sky intensities and the background model components,
respectively, that are adjusted by the fit.

%________________________________________________________________

\section{\label{results} Results}

\subsection{\label{results_mapping} Mapping}

\begin{figure*}
%\centerline{\epsfig{figure=aitoff_iter30.ps,width=13.5cm}}
%\centerline{\epsfig{figure=/users-data/weiden/SPI/PosCont/obs/All0019-0130-pub041210+vela/images/mrem2.ellipse_1.25_-0.75_10_6.CrabCygX1.ds1-f0.od.d.d.box5-5.bulge/aitoff_iter8.ps,bbllx=57pt,bblly=369pt,bburx=508pt,bbury=593pt,clip=,width=15.cm}}
%\centerline{\epsfig{figure=4046fig1.ps,bbllx=57pt,bblly=369pt,bburx=508pt,bbury=593pt,clip=,width=15.cm}}
\centerline{\epsfig{figure=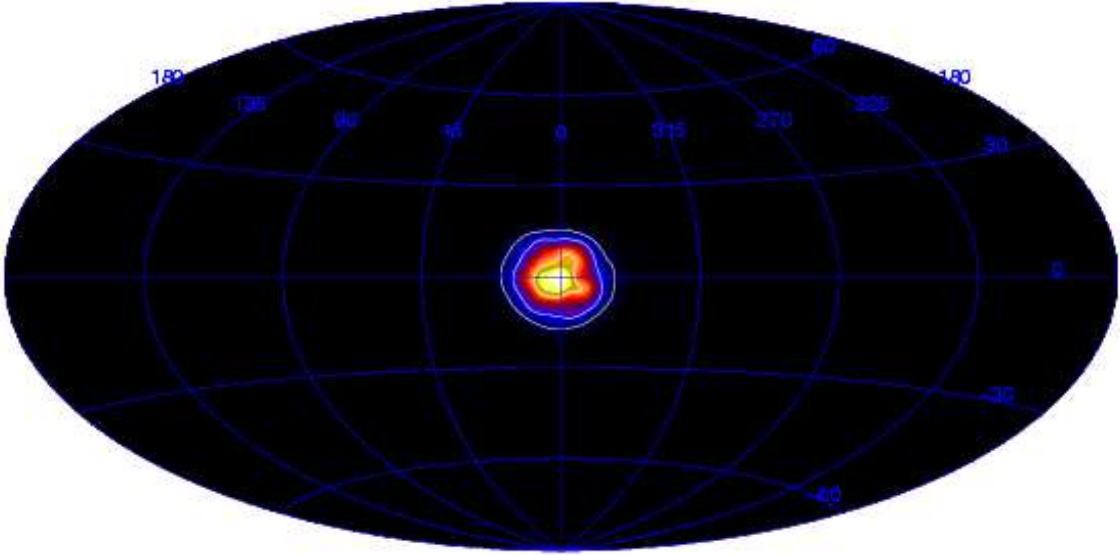,bbllx=20pt,bblly=282pt,bburx=472pt,bbury=510pt,clip=,width=15.cm}}
\caption{A Richardson-Lucy sky map of extended emission in the summed Ps analysis intervals (the combination of the intervals
410--430, 447-465, and 490--500~keV).
%The emission appears to be symmetric about the Galactic center, its
%centroid coincides well with the Galactic center. 
The contour levels indicate intensity levels of $10^{-2}$, $10^{-3}$,
and $10^{-4}$ ph~cm$^{-2}$~s$^{-1}$~sr$^{-1}$. Details are given in
the text.}
\label{fig1}
\end{figure*}

We employed an implementation of the Richardson-Lucy algorithm,
outlined in Sec.~\ref{data_analysis},
%\citep{Knoedlseder05} 
to obtain a model independent map of the extended emission in the
summed Ps analysis intervals (i.e.\ the events were first summed and
then analyzed, hence the total emission in the three energy intervals
is mapped without constraints on the relative fluxes in each of them;
spectroscopy of the emission is presented in
Sec.~\ref{spectral_analysis}).
%We employed an implementation of the Richardson-Lucy algorithm
%\citep{Knoedlseder05} to evaluate the source terms in fits of sky
%exposures (see Sec.~\ref{data_analysis}), providing us with a
%model-independent map of the extended emission in the combined Ps
%analysis intervals (i.e.\ the events were first summed and then analyzed).
%energy intervals 410--430, 447--465, and 490--500~keV.  
Since we are only interested in extended emission, the two point
sources Crab and Cygnus~X-1 (which are clearly seen due to their
strong continuum emission at these energies) were subtracted in the
image reconstruction process. In addition, to reduce noise artefacts,
the iterative corrections were smoothed during image reconstruction
with a $5^\circ \times 5^\circ$ boxcar average. The resulting sky map
after iteration 8 is presented in Fig.~\ref{fig1}. We selected this
iteration because at this point of the iterative mapping procedure the
fit quality corresponds approximately to that which we obtained when
fitting sky models to the data (see
Sec.~\ref{results_model_fitting}). This choice ensures that the
algorithm has not yet overfitted the data, which could result in
substantial image noise and image artefacts.
%However, simulations
%have demonstrated that faint, extended emission, as expected from the
%Galactic disk, may not yet be present in the map.
We note, however, that analyses of simulated observations show
that at this point in the iterations and with the limited amount of
data available, features such as the faint extended emission expected
from the Galactic plane
%may not yet have appeared.
would not necessarily be apparent, and indeed the Galactic plane
is not apparent in Fig.~\ref{fig1} although we detect it at a low
significance level in model fits (see
Sec.~\ref{results_model_fitting}). The simulations also show that
%(\dots after more iterations features which are not significant may
%appear due to overfitting\dots) 
with increasing iterations (in our analysis beyond iteration 8) faint
extended emission begins to appear, however, so do artefacts due to
overfitting, and a distinction in the image between actual signal and
noise becomes increasingly difficult.

The only prominent signal seen is
that from the Galactic bulge region. The surface brightness of any
emission from any other sky regions is much fainter. The emission
appears to be symmetric about the GC, and its centroid coincides well
with the GC. In short, the characteristics of the extended emission in
the three Ps analysis intervals agree well with those already
identified by us for the 511~keV line radiation. This agreement
suggests that the extended emission from the GC region in the Ps
analysis intervals is dominated by Ps continuum emission, with the
Galactic continuum emission being fainter (as will be demonstrated in
Sec.~\ref{spectral_analysis}). 
We note that these characteristics are consistent with
OSSE maps obtained for the Ps continuum emission
\citep[e.g.][]{Milne01b,Milne02}, in which the surface brightness is
highest in the bulge.

We also used the SPIROS algorithm
\citep{Skinner_Connell_03} to search for point sources in these three
energy bands. No significant sources in addition to the Crab and
Cygnus~X-1 were found outside the GC region. When imaging the GC
region with SPIROS, the observed emission formally is equally well
explained by six point sources. However, different sets of sources are
found in the different Ps analysis intervals. These sources do not
coincide with any known high-energy sources, including the four
highest energy sources reported by \citet{Bouchet05}. The spectrum of
these four sources drops steeply above about 300~keV, and since we are
analyzing rather narrow energy intervals above 400~keV the fact that
we do not yet detect them is not surprising. We therefore conclude
that the point sources found by us using SPIROS are all spurious,
resulting from SPIROS' attempt to account for
intrinsically diffuse emission with a set of point sources.

%{\bf ??? same sources in all energy bands ???}

\subsection{\label{results_model_fitting} Model fitting}

\begin{table}
%\caption[]{}
\caption{A summary of results from model fitting in the summed
Ps analysis intervals (the combination of the intervals 410--430,
447--465, and 490--500~keV). In the table, $\lambda$ is the maximum
likelihood ratio of the fit, $f_b$, $f_{\mathit{HI}}$, and
$f_{\mathit{CO}}$ are the fluxes in the bulge and the two different
disk components, respectively. The remaining symbols are defined in
the text. The error bars reflect the statistical $1\sigma$ confidence
interval for one degree of freedom.}
\label{table1}
\begin{center}
%     $$ 
%\begin{array}{p{0.5\linewidth}l}
%            \hline
%            \noalign{\smallskip}
%            Source      &  T / {[\mathrm{K}]} \\
%            \noalign{\smallskip}
%            \hline
%            \noalign{\smallskip}
%            Yorke 1979, Yorke 1980a & \leq 1700^{\mathrm{a}}     \\
%%           Yorke 1979, Yorke 1980a & \leq 1700             \\
%            Kr\"ugel 1971           & 1700 \leq T \leq 5000 \\
%            Cox \& Stewart 1969     & 5000 \leq             \\
%            \noalign{\smallskip}
%            \hline
%         \end{array}
%     $$ 
\begin{tabular}[t]{lc}
\hline 
\noalign{\smallskip}
Quantity & Value \\ \hline \hline
\noalign{\smallskip}
\multicolumn{2}{c}{Bulge only}\\ \hline
\noalign{\smallskip}
%$\lambda^\dagger$ & 494.5 \\
$\lambda$ & 494.5 \\
$l_0$ [$^\circ$] & $1.3 \pm 0.8$ \\
$b_0$ [$^\circ$] & $-0.8 \pm 0.6$ \\
FWHM$_l$ [$^\circ$] & $10.0^{+4.1}_{-1.9}$ \\
FWHM$_b$ [$^\circ$] & $6.6^{+1.4}_{-1.2}$ \\
%$\epsilon^\ddagger$ & $0.66^{+0.22}_{-0.20}$\\
$\epsilon$ & $0.66^{+0.22}_{-0.20}$\\
$f_b$ [ph/cm$^2$/s] & $1.20^{+0.14}_{-0.11} \times 10^{-3}$  \\
\noalign{\smallskip}
\hline
\noalign{\smallskip}
\multicolumn{2}{c}{Bulge + HI} \\ \hline
\noalign{\smallskip}
$\lambda$ & 502.2 \\
$l_0$ [$^\circ$] & $1.3 \pm 0.8$ \\
$b_0$ [$^\circ$] & $-0.9 \pm 0.6$ \\
FWHM$_l$ [$^\circ$] & $9.3^{+2.6}_{-1.5}$ \\
FWHM$_b$ [$^\circ$] & $6.8^{+1.5}_{-1.2}$ \\
$\epsilon$ & $0.71^{+0.26}_{-0.20}$\\
%$f_b$ [ph/cm$^2$/s] & $(1.16 \pm 0.11) \times 10^{-3}$  \\
$f_b$ [ph/cm$^2$/s] & $1.16^{+0.12}_{-0.10} \times 10^{-3}$  \\
%$f_{\mathit{HI}}$ [ph/cm$^2$/s] & $(4.18 \pm 1.52) \times 10^{-3}$ \\
$f_{\mathit{HI}}$ [ph/cm$^2$/s] & $4.19^{+1.52}_{-1.53} \times 10^{-3}$  \\
\noalign{\smallskip}
\hline
\noalign{\smallskip}
\multicolumn{2}{c}{Bulge + CO} \\ \hline
\noalign{\smallskip}
$\lambda$ & 510.5 \\
$l_0$ [$^\circ$] & $1.3 \pm 1.1$ \\
$b_0$ [$^\circ$] & $-1.2 \pm 1.0$ \\
FWHM$_l$ [$^\circ$] & $9.2^{+3.1}_{-2.1}$ \\
FWHM$_b$ [$^\circ$] & $7.8^{+3.2}_{-1.7}$ \\
$\epsilon$ & $0.85^{+0.44}_{-0.27}$\\
%$f_b$ [ph/cm$^2$/s] & $(0.86 \pm 0.14) \times 10^{-3}$  \\
$f_b$ [ph/cm$^2$/s] & $0.86^{+0.15}_{-0.13} \times 10^{-3}$  \\
%$f_{\mathit{CO}}$ [ph/cm$^2$/s] & $(1.97 \pm 0.48) \times 10^{-3}$ \\\hline
$f_{\mathit{CO}}$ [ph/cm$^2$/s] & $1.92^{+0.49}_{-0.48} \times
10^{-3}$ \\ \noalign{\smallskip} \hline
%\noalign{\smallskip} \hline \\
%\noalign{\smallskip}
%\multicolumn{2}{l}{$^\dagger$ Maximum likelihood ratio}\\
%\multicolumn{2}{l}{$^\ddagger$ Ellipticity $\epsilon = \mathrm{FWHM}_b /
%\mathrm{FWHM}_l$} \\  
%\noalign{\smallskip} \hline
\end{tabular}
\end{center}
\end{table}

A more quantitative approach for studying the Galactic distribution of
the observed extended emission is model fitting, which we performed using a
maximum likelihood multi-component fitting algorithm
\citep{Knoedlseder05} outlined in Sec.~\ref{data_analysis}.
% to evaluate the source terms in fits of sky exposures (see
%Sec.~\ref{data_analysis}). 

We first modelled the emission in the three summed Ps analysis
intervals\footnote{Results for the individual energy intervals are
consistent within statistical uncertainties.}
%three energy bands of interest
by an ellipsoidal distribution with a Gaussian radial profile and
determined the best-fit centroid location $(l_0, b_0)$ and extent in
Galactic longitude and latitude (FWHM$_l$, FWHM$_b$). We then combined
this Galactic bulge model with one of two models for emission from the
Galactic disk: both HI \citep{Dickey_Lockman90} and CO
\citep{Dame87} distributions are tracers of Galactic
matter and are believed to correlate with diffuse emission
\citep[cf.][]{Harris90, Kinzer99, Strong04}. The results of these fits
are summarized in Table~\ref{table1}.
%In all of these fits, steady
%point sources were included at the positions of the Crab and Cyg~X-1.
In each of these fits, the Crab and Cygnus~X-1 were included as steady
point sources whose intensities were fitted. When including the four
highest-energy sources reported by \citet{Bouchet05} the quality of
the fits is only slightly improved and the fit results do not change
significantly; therefore these point sources were excluded from the
final analysis.

%{\bf !!! fulx errors: asymmetric is better ???}

%{\bf ??? difference in centroids between Ps band and 511 line is
%marginally signifiant!!! but: Ps is not only Ps, but also Galactic
%continuum and point sources \dots ???}

As can be seen from Table~\ref{table1}, the centroid of the bulge
emission is the same within errors for all three models. There is
marginal evidence for a slight offset of the centroid from the GC, but
%an offset of this order can also result from  
it is of a magnitude that could easily result from the combined
effects of statistical and systematic biases in the background model
\citep[indeed, there is a similarly marginal, but opposite, offset of the
centroid in the 511~keV line emission; ][]{Knoedlseder05}. The extent
of the bulge emission, and its flux, do depend on the sky model. If
the extended emission is modelled by a bulge component only, then
there is marginal evidence for the bulge emission to be more extended
in longitude than in latitude (the ellipticity $\epsilon \equiv
\mathrm{FWHM}_b / \mathrm{FWHM}_l$ deviates by about $1.5\sigma$ from
unity). However, inclusion of a Galactic disk component improves the
fits, with the significances of the HI distribution
%\citep{Dickey_Lockman90} 
and of the CO distribution 
%\citep{Dame87}
being about $2.8\sigma$ and $4.0\sigma$, respectively, favouring the latter.
% with the CO distribution \citep{Dame87} being favoured over the
%HI distribution \citep{Dickey_Lockman90}. 
Another reason to adopt the CO distribution as the better disk model
of the two is the fact that the resulting total sky flux of about $(2.8
\pm 0.5) \times 10^{-3}$~ph~cm$^{-2}$~s$^{-1}$ agrees well with the
value of about $2.5 \times 10^{-3}$~ph~cm$^{-2}$~s$^{-1}$ determined
with SMM\footnote{The Gamma Ray Spectrometer on board the Solar
Maximum Mission \citep{Forrest80}.} in the Ps analysis
intervals, whereas the total bulge and HI disk model flux of $(5.4 \pm 1.5)
\times 10^{-3}$~ph~cm$^{-2}$~s$^{-1}$ is only marginally consistent
with the SMM spectrum of \citet{Harris90}.
%Evidence for ellipticity of the bulge, $\epsilon \equiv
%\mathrm{FWHM}_b / \mathrm{FWHM}_l$, is then marginal at best. 
%The bulge shape is consistent with symmetry, and bulge shape and extent
%are consistent with those found for the 511~keV line
%\citep{Knoedlseder05}. 

Inclusion of a Galactic disk component in the fits also renders
evidence for ellipticity of the bulge component insignificant. The
bulge shape is consistent with circular symmetry, with a
FWHM of about $8^\circ$, in agreement with our results for the 511~keV
line \citep{Knoedlseder05}. As is the case for the annihilation line,
the extent of the Ps continuum bulge emission is slightly larger than
that derived by \citet{Kinzer01} from OSSE observations. However, the
difference is not very significant, and it is possible that there is
bias in the OSSE analysis favouring a smaller bulge extent
\citep{Kinzer01}.

The fluxes that are attributed to the disk components exceed the bulge
flux by factors of 2--4 (see Table~\ref{table1}). However, since the
disk flux is distributed over a much larger sky region, the
corresponding surface brightness is much lower. The model fits
therefore confirm the mapping result: the intensity of extented
Galactic emission in the three Ps analysis intervals is brightest in
a region, several degrees across, around the GC.

%{\bf !!! TBD: re-scale fluxes from power-law slope 1.4 to 0 !!!}

\subsection{\label{spectral_analysis} Spectral analysis}

To assess the relative contributions of Ps continuum and Galactic
continuum emission to the emission mapped in Fig.~\ref{fig1}, we
performed a spectral analysis in six energy intervals: the three bands
used for the Ps continuum analysis supplemented by the energy
intervals 500--507, 507--515, and 515--565~keV. More detailed
spectroscopy including the line shape is presented in \citet{Jean06}.
%In each energy
%interval the instrumental background was modelled as described in
%Sec.~\ref{data_analysis}. Following our best sky model (see
%Table~\ref{table1}), and in agreement with our results on the sky
%distribution of the 511~keV line emission, we assumed for this
%analysis that the bulge emission is spherically symmetric about the GC
%with a Gaussian radial profile and a FWHM of $8^\circ$, and that the
%disk emission is traced by the line-of-sight integrated CO
%distribution. 
We want to emphasize that present uncertainties in the sky
distribution of the emission components, in particular for the
Galactic continuum emission, imply systematic uncertainties for this
kind of spectral analysis. Similarly, when comparing different
results, one has to keep in mind that there will be a systematic
difference between fluxes obtained under different assumptions for the
spatial distribution of the emission.

\subsubsection{\label{spectral_analysis_spi_simple} Simplistic SPI
spectroscopy}

A first, and rather simplistic, spectral analysis of the SPI result
for the diffuse emission from the GC region ($|l|, |b| \le 16^\circ$)
%obtained with our spatial model G8CO
%depicted in Figs.~\ref{fig2} and \ref{fig3} 
was performed by fitting
three spectral components: a line at 511~keV, the Ps continuum, and a
power law\footnote{The power law is defined as \\ $f_{\mbox{pl}}(E) = A
\cdot (E/300~keV)^{-\alpha}$} representing the Galactic
continuum. Before fitting, these models were convolved with the SPI
instrumental resolution.
Following our best sky model (``Bulge+CO'', see Table~\ref{table1}),
and in agreement with our results on the sky distribution of the
511~keV line emission, the SPI fluxes used in this fit were obtained
assuming that the bulge emission is circularly symmetric about
the GC with a Gaussian radial profile and a FWHM of $8^\circ$, and
assuming that the disk emission is traced by the line-of-sight
integrated CO distribution (hereafter: model G8CO).
%

%Following our best sky model (see
%Table~\ref{table1}), and in agreement with our results on the sky
%distribution of the 511~keV line emission, we assumed for this
%analysis that the bulge emission is spherically symmetric about the GC
%with a Gaussian radial profile and a FWHM of $8^\circ$, and that the
%disk emission is traced by the line-of-sight integrated CO
%distribution. 
The limited data available for this spectral analysis required that
some parameters of our spectral model be fixed or constrained.  The
centroid and width of the astrophysical positron annihilation line
were fixed at 511~keV and 2.5~keV FWHM \citep[c.f.][]{Lonjou04,
Churazov05}, respectively. In addition, the power-law photon index
$\alpha$ was fixed at 1.75 \citep[c.f.][]{Kinzer99, Bouchet05}. In
this first analysis,
%the instrumental resolution was assumed to be 2.1~keV FWHM at all
%energies, and a diagonal response was assumed, 
off-diagonal elements in the SPI response, which are small, were
neglected. When fitting the spectral models to the data, the models were
averaged over the energy intervals pertaining to each data point.
%, which is acceptable in view that our main goal is to
%investigate whether the Ps continuum is indeed the dominant emission
%component in the three Ps continuum analysis intervals that were the
%focus of our spatial analyses in Secs.~\ref{results_mapping} and
%\ref{results_model_fitting}. 
%A more detailed spectral analysis including the line shape is deferred
%to a future publication \citep{Jean06}.
A rigorous spectral analysis, using the full response matrix of SPI,
will be presented below. The advantage of this first analysis is its
simplicity, and the fact that its results can easily be displayed
graphically. 

\begin{figure}
%
%\centerline{\epsfig{figure=ellipse_0_0_8_8.6bands.CO.CrabCygX1.f0.od.d.d.compare_bulgedisk16_simplefit.ps,bbllx=82pt,bblly=374pt,bburx=516pt,bbury=640pt,clip=,width=8.5cm}}
%\centerline{\epsfig{figure=ellipse_0_0_8_8.6bands.CO.CrabCygX1.f0.od.d.d.compare_bulgedisk16_simplefit.ps,bbllx=82pt,bblly=402pt,bburx=516pt,bbury=695pt,clip=,width=8.5cm}}
\centerline{\epsfig{figure=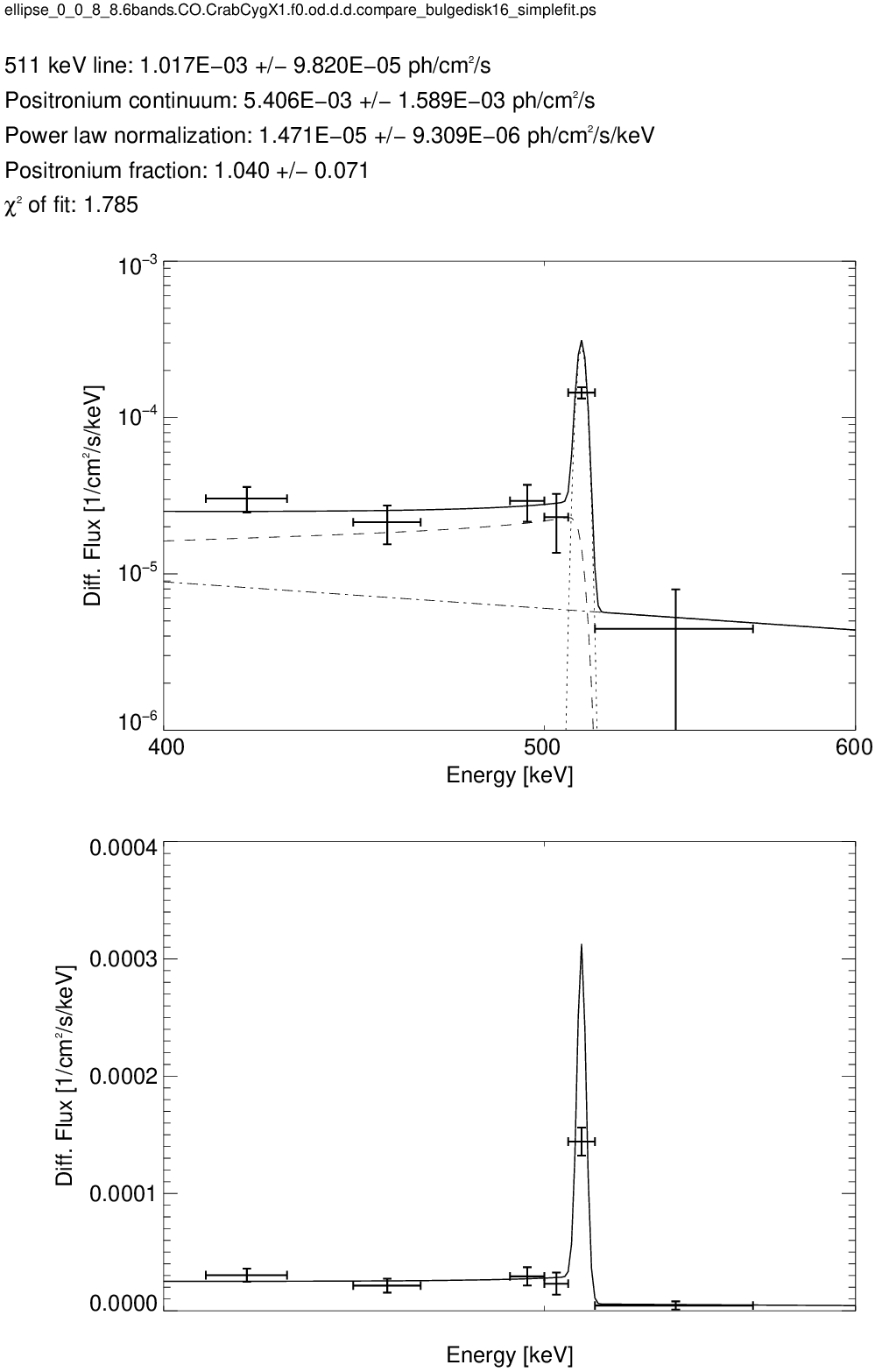,bbllx=82pt,bblly=402pt,bburx=516pt,bbury=695pt,clip=,width=8.5cm}}
%
%\caption{Comparison of SPI results on the diffuse emission from our
%Galaxy with results for the total emission from the Galaxy obtained
%with SMM data by \citet{Harris90} (top panel), and for the emission
%from the Galactic center region obtained with OSSE data by
%\citet{Kinzer99} (bottom panel).}
%
%\caption{Top panel: comparison of the SPI result for the diffuse
%emission from our Galaxy (data points) with results for the emission
%from the inner Galaxy obtained with SMM data by
%\citet{Harris90} (the solid and dashed lines represent their best fit
%model and its total uncertainty). \newline Bottom panel:
\caption{A fit of the SPI result for
the diffuse emission from the GC region ($|l|, |b| \le 16^\circ$)
obtained with a spatial model consisting of an $8^\circ$ FWHM Gaussian
bulge and a CO disk. In the fit a diagonal response was assumed. The
spectral components are: 511~keV line (dotted), Ps continuum (dashes),
and power-law continuum (dash-dots). The summed models are indicated
by the solid line. Details of the fitting procedure are given in the
text.}
\label{fig4}
\end{figure}

The best fit result is shown in Fig.~\ref{fig4}. 
We find a 511~keV line flux of $(1.02 \pm 0.10) \times
10^{-3}$~ph~cm$^{-2}$~s$^{-1}$, and a Ps continuum flux of about $(5.41 \pm
1.59) \times 10^{-3}$~ph~cm$^{-2}$~s$^{-1}$. 
%, in good agreement for the Galactic
%bulge flux determined by \citet{Knoedlseder05} in a dedicated
%annihilation line analysis. The Ps continuum flux is about $(5.31 \pm
%1.59) \times 10^{-3}$~ph~cm$^{-2}$~s$^{-1}$, 
From these fluxes, and using simple error propagation, we obtain a
value of $f_{Ps}$ of $1.04 \pm 0.07$ for the Ps fraction as defined by
\citet{Brown_Leventhal87}.
%yielding a Ps fraction $f_{Ps}$ of $1.04 \pm 0.07$.
%\newline
%{\bf TBD ??? explain in paper where the value and error comes from; simple
%error propagation using Brown and Leventhal defintion; or explain this
%only in reference to referee ???}
%\newline
%, consistent with previous results. 
The power law is not significantly detected with a normalization of $(1.47
\pm 0.93) \times 10^{-5}$~ph~cm$^{-2}$~s$^{-1}$~keV$^{-1}$. The
$\chi^2$ of the fit is about 1.8 for 3 degrees of freedom. Based
on the best fit parameter values, the Ps continuum emission accounts for
about 71\% of the total emission in the three Ps continuum analysis
intervals mapped in Fig.~\ref{fig1}.
%{\bf ???? Despite the simplicity of this approach, the results are
%consistent with a detailed analysis described in the following. ???}

\subsubsection{\label{spectral_analysis_spi_rigorous} Rigorous SPI
spectroscopy}

To perform a rigorous spectral analysis of the emission from the
Galactic bulge, we fitted 
%As an alternative approach, we attempted to fit 
the same spectral model -- a line at 511~keV, the Ps continuum, and a
power law
%\footnote{In XSPEC, a power law is defined by \\ $pl(E) =
%A \cdot (E/1~keV)^{-\alpha}$} 
-- to a grid of points centred on the
GC using the INTEGRAL/SPI specific capabilities of the
XSPEC (ver 12) analysis package
\citep[see e.g.][]{Dorman_Arnaud01}, which allowed us to take into
account the full spectral and spatial response of the instrument. For
a detailed description of this methodology the reader is referred to
\citet{Shrader04}; here we will just briefly describe the basic
procedure followed. This is the first time that XSPEC has been applied
to spectroscopy of an extended sky source observed with the SPI
instrument.
%
%We performed a spectral analysis of the SPI result employing our best
%fit spatial model, consisting of an ellipsoidal bulge component and a
%CO distribution disk component. For each of these two spatial
%components, a three component spectral model, consisting of Ps
%continuum, 511~keV line, and Galactic continuum, was assumed. This
%model was fitted to a grid of points centered on the Galactic bulge
%using the INTEGRAL/SPI specific capabilities of the XSPEC (ver 12)
%analysis package \citep[see e.g.][]{Dorman_Arnaud01}. For a detailed
%description of this methodology the reader is referred to
%\citet{Shrader04}; here we will just briefly describe the basic
%procedure followed. 
As an aside note, we wish to warn the interested reader that we found
the original Ps continuum model in XSPEC, {\tt POSM}, to be
incorrectly implemented. We developed and tested a new implementation
of the \citet{Ore_Powell49} spectral shape of Ps continuum emission,
which will be included in subsequent releases of XSPEC.

The data selected for this portion of our analysis comprise a
subset of the total data presented in this
paper. Observations were selected for inclusion in our spectral
fitting when the SPI telescope axis was aligned with the GC to within
an angular offset 
%of the nominal fully-coded SPI field-of-view 
of $16^\circ$ (the extent of the nominal fully-coded SPI
field-of-view). This resulted in a total of about 750 spacecraft
pointings (Science Windows), totalling $\sim$1.7~Ms of live time, being used
in this analysis.  

The full SPI instrument response, including
diagonal plus off-diagonal matrix elements, was then computed,
according to the methodologies described in \cite{Sturner03}, for each
SPI detector for each selected instrument pointing for each of our
grid points spatially sampling the bulge region. Specifically, we computed
the response for a 21-point raster at $(l,b) = (0^\circ,0^\circ),
(\pm4^\circ,0^\circ), (0^\circ,\pm4^\circ), (\pm8^\circ,0^\circ),
(0^\circ,\pm8^\circ)$, $(\pm4^\circ,\pm4^\circ),
(\pm8^\circ,\pm4^\circ), (\pm4^\circ,\pm8^\circ)$.
%
%%The response for each detector for each
%%pointing was computed for each , and
%%according to the methodologies described in \cite{Sturner03}. 
%%The same energy binning scheme defined in Secs.~\ref{results_mapping} and
%%\ref{spectral_analysis} was employed. 
%The data were background subtracted in detector count space using the
%solution for the background obtained from the model fit results
%described in Sec.~\ref{results_model_fitting}.
%%as described in
%%Sec.~\ref{data_analysis}, although 
%We note that the capability to simultaneously solve for the background
%and source content of a given observation exists within the XSPEC (ver
%12) implementation \citep{Shrader04}. In practice however, that has
%proven difficult.

The data were then simultaneously fitted to the physical model
described above -- 511~keV line, Ps continuum, and power law -- and
the 3-component background model described in Sec.~\ref{data_analysis}.
%(the fourth component was fixed at zero). 
The background model in this
case was parameterized so that small ($\pm$10\%) variations were
allowed for the normalization terms of each component in each energy
interval, using the results of model fits (as decribed in
Sec.~\ref{results_model_fitting}) to initialize the background model
parameters. In practice we found that the background modelling worked
quite well, with the best fit solutions typically corresponding to
normalization terms within $\pm$1\% of unity.

We then made the assumption that the net flux consists of additive
contributions from the two spatial models discussed in
Sec.~\ref{spectral_analysis_spi_simple}, i.e.\ the Gaussian and CO
distributions of spatial model G8CO. The spectral model was then
applied to the SPI instrument response function twice at each spatial
raster, with a normalized, relative, weighting factor based on both
the 
%ellipsoidal
Gaussian and the CO distributions.
%The quoted flux determinations are the
%integration of the derived model over the covered energy range and
%summed over the two spatial distributions {\bf ????????}. 
This leads to a data space which scales as: (number of SPI pointings)
$\times$ (number of detectors) $\times$ (number of spectral
channels). This number is then multiplied by (number of spatial
rasters) $\times$ (2 spatial distribution models) to give the number
of individual response matrices applied to the spectral model for the
$\chi^2$ minimization problem. This leads to $\sim 750 \times 19
\times 6 \times 21 \times 2 \sim 3.6 \times 10^6$ folded-model
calculations per iteration step of the $\chi^2$ minimization
procedure. Specifically, we used the XSPEC ``FLUX'' command and the
best fit parameters of each individual model component to integrate
over the covered energy range.

%A $\chi^2$ minimization procedure was performed by independently
%varying various model components. The line centroid and width were
%fixed at 511~keV and 1.06~keV (i.e.\ 2.5~keV FWHM) respectively. We
%constrained the powerlaw photon index to approximately $1.75\pm0.10$
%\citep[c.f.][]{Kinzer99, Bouchet05}, and the amplitude to within a
%factor of $\sim$3 to that invoked in section X {\bf ??? TBD
%???}. Otherwise, the model parameters were allowed to vary freely, and
%separately from grid point to grid point (or "source"). The quoted
%results are an average over the composite fit.

The parameter space was constrained as follows. The centroid and width
of the positron annihilation line were fixed at 511~keV and 2.5~keV
FWHM, respectively, as in our first analysis (see
Sec.~\ref{spectral_analysis_spi_simple}). We fixed the power-law
photon index $\alpha$ to a value of 1.75, but allowed the amplitude to
vary by about a factor of 4 relative to that obtained in our first
analysis described above. Otherwise, the model parameters --
specifically the Ps continuum and Gaussian line normalization terms --
were allowed to vary freely in the $\chi^2$ minimization. These two
normalization terms were varied separately with respect to the two
spatial distributions, but linked from grid point to grid point within
a given spatial model. This leads to 6 free physical model parameters
(3 normalizations for each of the 2 spatial models), in addition to
the 18 background model parameters (3 parameters in each of the 6
energy intervals) for the overall fit.

%%The results we obtained are consistent within statistics to the model
%%depiction presented in Fig.~\ref{fig4} {\bf ??? check !!!}. 
%We obtained a Ps continuum normalization of $3.92 \pm {\rm ???} \times
%10^{-3}$~ph~cm$^{-2}$~s$^{-1}$. Combined with the inferred
%gaussian line component normalization of $(9.26 \pm 0.09) \times
%10^{-4}$~ph~cm$^{-2}$~s$^{-1}$ we obtain a Ps fraction of $f_{Ps} =
%0.98\pm0.10$. The normalization of the power law component is $0.09
%\pm$ {\bf ???}, its index is $1.70 \pm$ {\bf ???}. 
%%
%%Our spectral
%%results are consistent with earlier measurements of the annihilation
%%radiation from the GC region
%%\citep[e.g.][]{Harris98, Milne00, Kinzer01, Churazov05,
%%Knoedlseder05}. 
%%The normalization of the power law component is {\bf
%%??? TBD ???}~ph~cm$^{-2}$~s$^{-1}$~keV$^{-1}$, in agreement with
%%\citet{Kinzer99}. The three spectral components are shown in
%%Fig.~{\bf ???}.

%We obtained a Ps continuum normalization of $(3.79 \pm 0.70) \times
%10^{-3}$~ph~cm$^{-2}$~s$^{-1}$. Combined with the inferred Gaussian
%line component normalization of $(9.25 \pm 0.58) \times
%10^{-4}$~ph~cm$^{-2}$~s$^{-1}$ we obtain a Ps fraction of $f_{Ps} =
%0.975\pm0.095$. The normalization of the power-law component is $0.089
%\pm0.007$ {\bf !!! TBD: units !!!}, its index is
%$1.70^{+0.10}_{-0.05}$. The background normalization terms, as noted,
%were within 1\% of unity. The $\chi^2_{\nu}$ values obtained was
%of order unity; for example, in a 0.2 Msec subset of the data a
%$\chi^2$ per DoF of 12084/11987 was achieved.

We obtained a Ps continuum normalization of $(3.11 \pm 0.56) \times
10^{-3}$~ph~cm$^{-2}$~s$^{-1}$. Combined with the inferred Gaussian
line component normalization of $(9.35 \pm 0.54) \times
10^{-4}$~ph~cm$^{-2}$~s$^{-1}$ we obtain a Ps fraction of $f_{Ps} =
0.92\pm0.09$. 
%
%{\bf We note that the uncertainties on the individual
%inferred parameters of the spectral model were derived using the
%covariance matrix. In particular, since this matrix can have positive
%and/or negative terms, the uncertainty of a combined parameter such as
%the Ps fraction can be less than the summation
%in quadrature of combined set or parameters}. 
%
The normalization of the
power-law component, rescaling the XSPEC result to the
power-law function defined in footnote~6, is
%$(8.2_{-2.7}^{+3.6})\times10^{-2}$~ph~cm$^{-2}$~s$^{-1}$~keV$^{-1}$. 
$(3.79_{-1.25}^{+1.66})\times10^{-6}$~ph~cm$^{-2}$~s$^{-1}$~keV$^{-1}$.
%\newline
%{\bf !!! TBD: ??? why the difference between simplistic fit and XSPEC fit,
%error bars ``touch''\dots ???}
%\newline
Thus over the range of our data, the power-law component contributes
approximately 6\% (and possibly as much as 14\%) of the continuum flux
based on our model fitting.  If we allow the power-law index to vary
freely, the resulting power-law flux remains within the confidence
interval above; hence our conclusion regarding the flux contribution
of the power-law component is robust.  The background normalization
terms, as noted, were within 1\% of unity. The $\chi^2_{\nu}$ value
obtained was of order unity; specifically, using the full 1.65~Ms of
the data selected for this analysis, a $\chi^2$ per degree of freedom
of $99065.1/86289\approx1.15$ was achieved.
The uncertainties for a given parameter, specficially the line and Ps
continuum fluxes and the power-law normalization, were derived
by varying the parameter within its allowed range. At each step, the
other free parameters are allowed to vary until the fit statistic is
minimized, determining the 1$\sigma$ confidence region for each
parameter (specifically, this is accomplished using the ``ERROR''
procedure of XSPEC v12). 
%For the Ps fraction, the uncertainty was
%estimated using the resulting line and continuum uncertainties, plus
%the cross-term of the covariance matrix resulting from our $\chi^2$
%minimization procedure.
We note that the uncertainty in the Ps fraction includes both the
variances and the covariances of the 511~keV line and Ps continuum
fluxes in the variance-covariance matrix of the fit.

Clearly, given the nature of the data and parameter spaces of the
problem, it is difficult to present a graphical representation of the
results. A procedure commonly used within XSPEC is to ``group'' the
data for plotting purposes, which produces what is essentially an
averaged representation of the data and model convolved with the
response. In this application, however, this is not a viable option 
%in our case 
as we would be averaging over some $\sim10^6$
spectra/responses/model calculations. We have thus chosen not to
include a plot of the XSPEC results.

%{\bf !!! following: TBD !!!}
%
%Our results on the annihilation radiation from the GC region are
%consistent with earlier measurements of the 511~keV line flux
%\citep{Harris98, Milne00, Knoedlseder05} and of the Ps
%continuum flux and the Ps fraction \citep{Harris98, Kinzer01,
%Churazov05}.

%{\bf Our result for the Ps fraction for the GC region agrees well with
%previous determinations by \citet{Harris98} using TGRS ($f_{P_S} =
%0.94 \pm 0.04)$, by \citet{Kinzer01} using OSSE ($f_{P_S} =
%0.93 \pm 0.04)$, and by \citet{Churazov05} ($f_{P_S} =
%0.96 \pm 0.06)$. }
%{\bf ??? flux ???}.

%It follows from this analysis that a flux of about $6.4 \times
%10^{-4}$~ph~cm$^{-2}$~s$^{-1}$ can be attributed to Ps continuum in
%the three Ps continuum analysis intervals, compared to a flux of about
%$1.4 \times 10^{-4}$~ph~cm$^{-2}$~s$^{-1}$ due to the Galactic
%continuum. 
%{\bf ??? line contribution ???}.

It follows from our spectral analysis of the Galactic bulge emission
that at least 76\% (the best value is 85\%), of the observed flux in
the summed Ps analysis intervals can be attributed
to Ps continuum emission -- which therefore dominates the emission
from the GC region.
The map in Fig.~\ref{fig1} is dominated by emission from the central
regions of our Galaxy. 
%Figure~\ref{fig3} clearly shows that this emission is dominated by Ps
%continuum emission. 
Hence we can conclude that Fig.~\ref{fig1} provides a view of
the sky distribution of Ps continuum emission.

\subsubsection{\label{spectral_analysis_compare_spi} Comparison to
SPI results}

We note that our 511~keV line flux is somewhat lower than the previous
determination of $(1.05 \pm 0.06) \times
10^{-3}$~ph~cm$^{-2}$~s$^{-1}$ for the Galactic bulge by
\citet{Knoedlseder05} at the $\sim2\sigma$ level (the uncertainties of
other earlier SPI measurements, e.g.\ those by \citet{Jean03a} or
\citet{Weidenspointner04}, are too large to reach a definitive
conclusion), and that the 511~keV line, the Ps continuum, and the
Galactic continuum
fluxes from our XSPEC analysis are lower than those obtained in our
more simplistic approach described above.
%, and particularly the Ps
%continuum flux are discrepant {\bf ????} with the previous
%determinations of Jean et al 20XX (line), and Knoedlseder et al. 20XX
%(Ps continuum).} The discrepancies are within about 1- and 2-sigma
%respectively, but in both cases, our results are lower. 
We suggest that this may be due to the fact that our XSPEC analysis
included the off-diagonal response matrix elements, while the previous
analyses used a diagonal approximation. Roughly speaking, the reason
for a lower flux in our case is that with more matrix elements a
smaller model flux is required to match the ``model counts'' with the
actual detector counts in the $\chi^2$ minimization procedure. We find
for example, that for a Crab-like power-law model applied to the SPI
full response leads to a $\sim8$\% lower flux than for the same model
applied to a diagonalized version of the response. Indeed, for the
actual Crab data used in determining the SPI calibration, the
discrepancy between flux determinations made with SPIROS
\citep[diagonal approximation; ][]{Skinner_Connell_03} and XSPEC was
initially about 15\% and in a similar sense (XSPEC flux was
lower). Subsequent efforts were made to calibrate the SPIROS method
for better agreement \citep{Sizun04}.  Finally, we note that the
effects of the off-diagonal terms are even more pronounced for the
case of a positively sloped spectrum, such as,
%the red-wing of a broad line feature or, 
notably, the expected Ps continuum form.

Our result for the Ps fraction of $f_{Ps} = 0.92\pm0.09$ agrees very
well with an earlier result of $f_{P_S} = 0.96 \pm 0.06$ derived by
\citet{Churazov05} for the Galactic bulge using SPI data.

We also note that \citet{Strong05} concluded, as we do, that the
emission from the Galactic bulge is dominated by Ps continuum
emission. Their results for $|l|, |b| \le 10^\circ$ imply that about
81\% of the emission in the Ps analysis intervals is due to Ps
continuum emission, in excellent agreement to our result.

\subsubsection{\label{spectral_analysis_compare_osse} Comparison to
OSSE results}

%Comparisons of SPI results for the diffuse emission from our Galaxy
%with results for the total emission from the inner Galaxy obtained
%from SMM data by \citet{Harris90} (which are not essentially
%different from the emission from the whole Galaxy), and for the
%total emission from the GC region obtained with OSSE data by
%\citet{Kinzer99}, are shown in Figs.~\ref{fig2} and \ref{fig3},
%respectively. 

The comparison of the astrophysical emission models
determined with other instruments with our SPI results was performed
by first convolving the spectral models with the SPI instrumental resolution,
and then by averaging the convolved models over the energy ranges of
the SPI data points. 
%
%Following our best sky model (``Bulge+CO'', see Table~\ref{table1}),
%and in agreement with our results on the sky distribution of the
%511~keV line emission, the SPI results used in the comparisons were
%obtained assuming that the bulge emission is spherically symmetric
%about the GC with a Gaussian radial profile and a FWHM of $8^\circ$,
%and assuming that the disk emission is traced by the line-of-sight
%integrated CO distribution (hereafter: model G8CO).
%
As in our spectroscopic analyses described above, the SPI results used in
these comparisons were obtained with spatial model G8CO.
It is worth noting that \citet{Harris90} had assumed in their SMM
analysis that the Galactic gamma-ray emission follows the CO
line-of-sight distribution of
\citet{Dame87}, and that \citet{Kinzer99} had concluded from their
OSSE analysis that the longitude distribution of the Galactic
continuum emission follows that of CO, albeit with a somewhat broader
distribution in latitude.
To make comparisons with OSSE results, we consider separately the form
of the spectrum, the annihilation flux, and the Galactic continuum
emission.

%The comparison to OSSE results is done in two steps. First, we will
%compare to the shape of the total spectrum determined for the GC
%center region by \citet{Kinzer99}, then we will compare to the
%annihilation flux from the inner square radian of our Galaxy
%determined by \citet{Kinzer01} {\bf and to the Galactic continuum flux from
%the GC region determined by \citet{Purcell96} and \citet{Kinzer99}. 
%%An absolute comparison with the \citet{Kinzer99} result is more
%%difficult, because the latter pertains to the total flux seen in
%%OSSE's diamond-shaped field-of-view.
%We would like to emphasize that an absolute comparison of flux levels
%is more difficult for results obtained by \citet{Purcell96} and
%\citet{Kinzer99} than for those obtained by \citet{Kinzer01} since the
%former two pertain to the flux in OSSE's diamond-shaped field-of-view
%rather than to a more well-defined sky region.
%%, thereby rendering the comparion difficult.
%}
%%
%%In addition,
%%\citet{Kinzer99} derived their result under a different assumption for
%%the spatial distribution of the emission (namely that the spatial
%%distributions of the Ps continuum and the Galactic continuum emission
%%are identical) than we, resulting in different systematic biases in
%%the flux determinations.  
%%
%%Nevertheless, the shapes of the SPI and OSSE spectra
%%can still be compared, which is interesting as it is the spectrum of
%%the emission from the Galactic bulge region that is relevant for
%%identifying the dominant emission component in the map of
%%Fig.~\ref{fig1}.

\begin{figure}
%
%\vspace*{2ex}
%
%\centerline{\epsfig{figure=ellipse_0_0_8_8.6bands.CO.CrabCygX1.f0.od.d.d.compare_bulgedisk20_special.ps,bbllx=82pt,bblly=402pt,bburx=512pt,bbury=695pt,clip=,width=8.5cm}}
%\centerline{\epsfig{figure=ellipse_0_0_8_8.6bands.CO.CrabCygX1.f0.od.d.d.compare_bulgedisk16.ps,bbllx=82pt,bblly=402pt,bburx=516pt,bbury=695pt,clip=,width=8.5cm}}
\centerline{\epsfig{figure=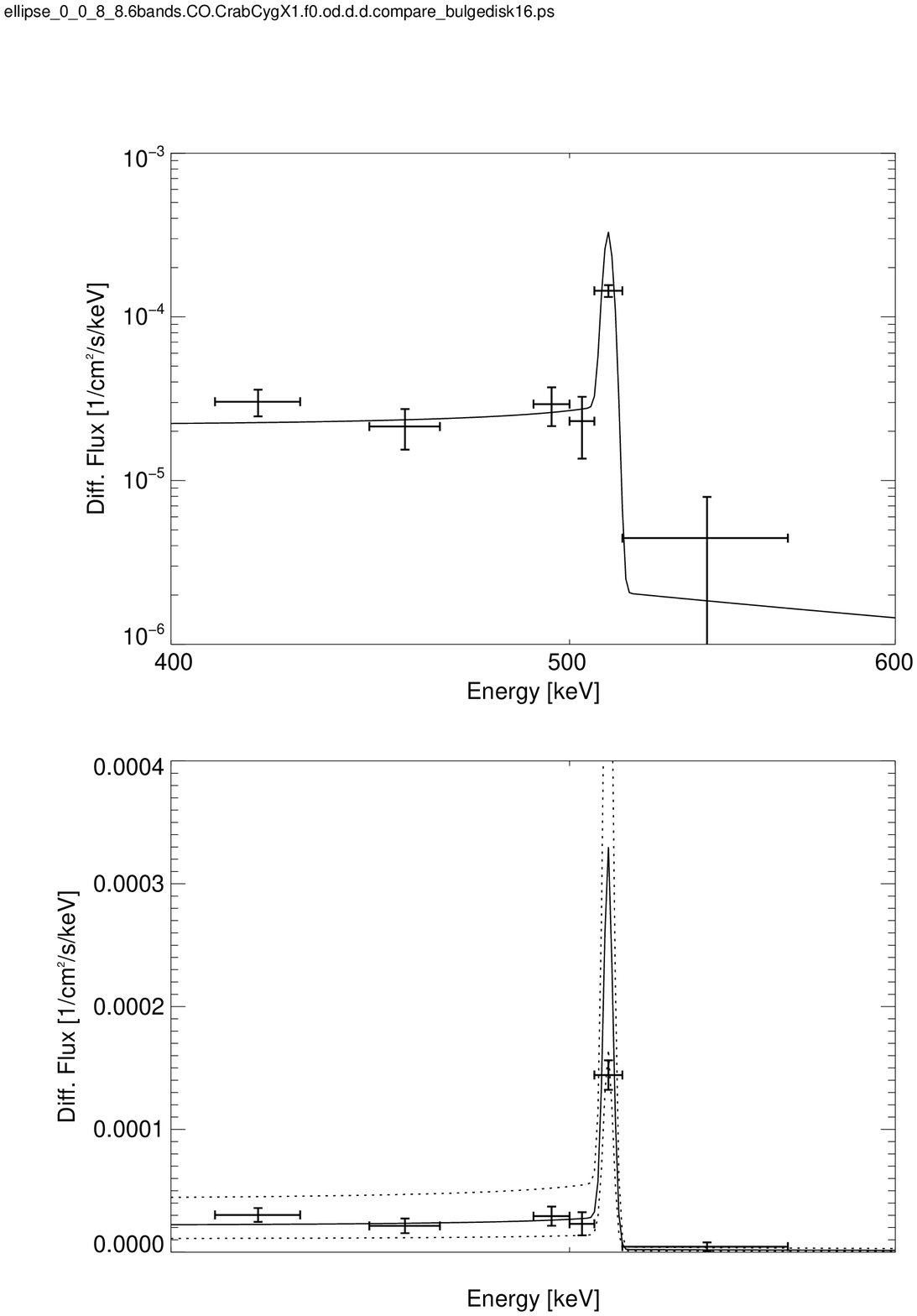,bbllx=82pt,bblly=402pt,bburx=516pt,bbury=695pt,clip=,width=8.5cm}}
%
%\vspace*{2ex}
%
\caption{A fit of the GC spectrum obtained from OSSE data by
\citet{Kinzer99} to the SPI result for the diffuse emission from the GC
region ($|l|, |b| \le 16^\circ$) obtained with a spatial model
consisting of an $8^\circ$ FWHM Gaussian bulge and a CO disk.
Details are given in the text.}
\label{fig3}
\end{figure}

A comparison of the Galactic spectrum seen by SPI in a box
%of $40^\circ \times 40^\circ$ centred on the GC
defined by $|l|, |b| \le 16^\circ$ for our spatial model G8CO
with the \citet{Kinzer99} OSSE best fit model is depicted in
Fig.~\ref{fig3}.
% convolved with the SPI instrumental resolution. 
In this comparison of spectral forms, the normalization of the OSSE
model was obtained by a fit to the SPI data.
% (again, in each bin the model was averaged before fitting). 
We find a normalization factor of about 0.6, and a $\chi^2$ value of
about 3.0 for 5 degrees of freedom. We conclude that the shape of the
total SPI spectrum agrees well with that of the total OSSE
spectrum for the GC region. 
%{\bf ???? As will be demonstrated below,
%the SPI fluxes for the positron annihilation emission are in agreement
%with those derived with OSSE ??? TBD: for what region ??? by
%\citet{Kinzer01}.}
%
%As will be shown below, also our fluxes for the annihilation radiation from
%within the central square radian around the GC are consistent with
%OSSE results obtained by \citet{Kinzer01}.
%
%
%The dashed and dash-dotted lines indicate the Ps continuum and the
%Galactic continuum spectrum (for which a power-law index of -1.75 was
%assumed) as derived from a fit to the SPI data, neglecting
%off-diagonal elements in the SPI response (whose effect is small). The
%width of the 511~keV line was assumed to be 2.7~keV FWHM {\bf ??? TBD
%???}. Our preliminary result for the positronium fraction is $1.02 \pm
%0.09$, consistent with previous findings {\bf !!! TBD: references !!!
%update with XSPEC result ?!?!?}. {\bf !!! TBD: absolute SPI fluxes
%!!!}. A more detailed spectral analysis including the line shape is
%deferred to a future publication \citep{Jean06}.
%
%The dashed and dash-dotted lines indicate the Ps continuum and the
%Galactic continuum spectrum as derived from a fit to the SPI data as
%described below. A more detailed spectral analysis including the line
%shape is deferred to a future publication \citep{Jean06}.

%To compare to the total annihilation radiation fluxes (i.e.\ the sum of the
%annihilation line and the Ps continuum fluxes) from within the central
%square radian around the GC that \citet{Kinzer01} obtained from a
%dedicated annihilation radiation analysis using OSSE data, we repeated
%the simplistic spectral analysis of
%Sec.~\ref{spectral_analysis_spi_simple} in $|l|, |b| \le 29^\circ$.

For comparison with the annihilation radiation analysis in the
central square radian around the GC by \citet{Kinzer01} using OSSE
data, we repeated the simplistic spectral analysis of
Sec.~\ref{spectral_analysis_spi_simple} in $|l|, |b| \le 29^\circ$
using spatial model G8CO.
%
%The same simplistic spectral analysis was repeated with spatial model
%G8CO for the flux from within the central square radian around the GC
%in order to compare with the dedicated annihilation radiation analysis
%that \citet{Kinzer01} performed using OSSE data. 
\citet{Kinzer01} obtained total annihilation radiation fluxes (i.e.\
the sum of the annihilation line and the Ps continuum fluxes) of about
$(6-11)\times10^{-3}$~ph~cm$^{-2}$~s$^{-1}$ when fitting nova
ellipsoid or Gaussian bulge models combined with different thin disk
models, which, of all their models, are the most similar to our G8CO
model. The SPI total annihilation flux is about
$7.5\times10^{-3}$~ph~cm$^{-2}$~s$^{-1}$, in very good agreement with
%\citet{Kinzer01}. 
their results.
%Our simplistic analysis does not allow for an
%accurate determination of the Ps fraction (see below), which is
%overestimated, so it is not
%surprising that our 511~keV line
%
Our measurement of the Ps fraction is also in very good
agreement with 
%\citet{Kinzer01} result for the Galactic bulge of $f_{P_S} = 0.93 \pm
%0.04$.
their value for the Galactic bulge of $f_{P_S} = 0.93 \pm
0.04$.

For the Galactic continuum emission comparison we computed the
flux in the 410--500~keV band, that is most important in our analysis.
Comparison with both the \citet{Kinzer01} and the \citet{Purcell96}
analyses of OSSE 
%results 
observations shows good agreement. Approximating the
cosmic-ray interaction model used in the \citet{Kinzer01} analysis by
a power law with photon index $1.75$, we find that their results
correspond to a flux of about $3.4 \times
10^{-4}$~ph~cm$^{-2}$~s$^{-1}$~rad$^{-1}$,
%about $7.7 \times
%10^{-6}$~ph~cm$^{-2}$~s$^{-1}$~keV$^{-1}$~rad$^{-1}$ 
%at 300~keV,
whereas assuming that our results refer to a $32^\circ$ range in
Galactic longitude we find $3.0^{+1.3}_{-1.0} \times
10^{-4}$~ph~cm$^{-2}$~s$^{-1}$~rad$^{-1}$. The \citet{Purcell96}
analysis implies a flux of about $3.6 \times
10^{-4}$~ph~cm$^{-2}$~s$^{-1}$~rad$^{-1}$, however, their power-law
index of $2.5$ is steeper than our result, which we think is due to
the fact that they used data
dominated by energies below a few hundred keV.

\subsubsection{\label{spectral_analysis_compare_smm} Comparison to
SMM results}

\begin{figure}
%
%\centerline{\epsfig{figure=ellipse_0_0_8_8.6bands.CO.CrabCygX1.f0.od.d.d.compare_bulgedisk.ps,bbllx=82pt,bblly=402pt,bburx=516pt,bbury=695pt,clip=,width=8.5cm}}
\centerline{\epsfig{figure=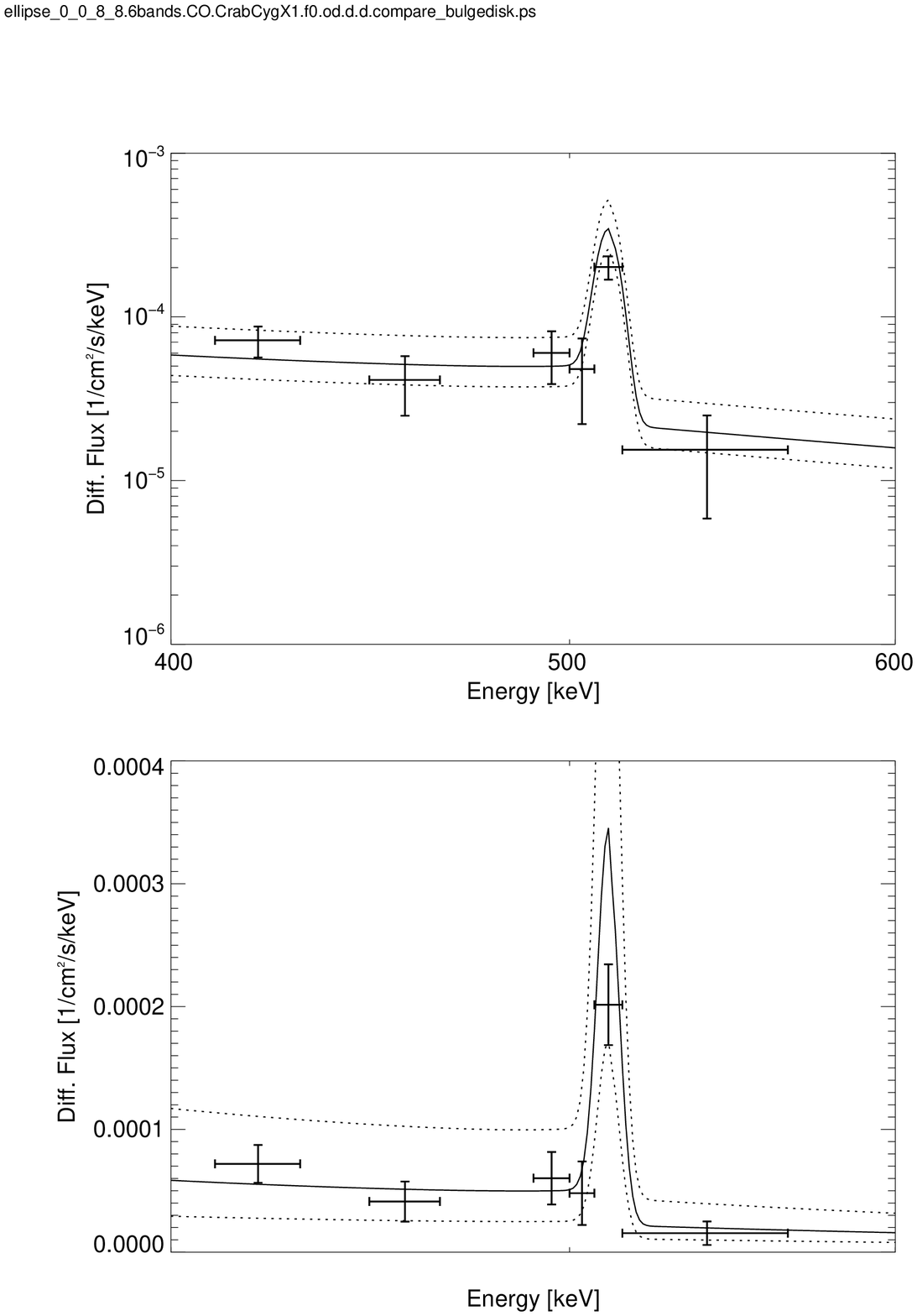,bbllx=82pt,bblly=402pt,bburx=516pt,bbury=695pt,clip=,width=8.5cm}}
%
%\centerline{\epsfig{figure=ellipse_0_0_8_8.6bands.CO.CrabCygX1.f0.od.d.d.compare_bulgedisk20.ps,bbllx=82pt,bblly=402pt,bburx=512pt,bbury=714pt,clip=,width=8cm}}
\caption{Comparison of the SPI result for the diffuse
emission from our Galaxy obtained with a spatial model consisting of
an $8^\circ$ FWHM Gaussian bulge and a CO disk (data points) with the
best fit model for the emission from the inner Galaxy obtained with
SMM data by \citet{Harris90} (the solid and dashed lines represent
their best fit model and its total uncertainty).
%, convolved with the SPI instrumental resolution.
Details are given in the text.
}
\label{fig2}
\end{figure}

A comparion of SPI results for the diffuse emission from our Galaxy
obtained with spatial model G8CO with results for the total emission
from the inner Galaxy obtained from SMM data by \citet{Harris90}
(which are not essentially different from the emission from the whole
Galaxy) is given in Fig.~\ref{fig2}. As can be seen, the SPI results
agree well with the best fit SMM model.
%(convolved with the SPI instrumental resolution), 
%as can be seen in Fig.~\ref{fig2}. 
Except for the 511~keV line
interval our fluxes agree within their statistical $1\sigma$ errors
with the \citet{Harris90} best fit model. In the 511~keV line region
our flux falls about $2\sigma$ short of the SMM result. Formally, a
comparison of the SMM model to the SPI data points yields a $\chi^2$
value of about 8.6 for 6 degrees of freedom
% (in each bin, the model was averaged before comparing to the SPI)
(the model amplitude was not fitted). 
%, but still agrees within the total uncertainty of the latter.
% 
%Our 511~keV line flux of about $(1.4 \pm 0.3) \times
%10^{-3}$~ph~cm$^{-2}$~s$^{-1}$ {\bf ??? TBD ???} is consistent with
%the total 511~keV line flux from our Galaxy as determined by us
%\citep{Knoedlseder05} in a dedicated line analysis.
%
% (the total uncertainty in the SMM flux is indicated by
%the dotted lines, SPI errors are statistical only). 
%{\bf !!! TBD: absolute SPI fluxes, compare to SMM fluxes !!!}

%________________________________________________________________

\section{\label{summary} Summary and conclusion}

We studied the sky distribution of Ps continuum emission.  
%The most important finding of this work is that the
The extent and shape of extended emission (consisting of Ps continuum
and Galactic continuum) in the Ps analysis intervals
%three energy bands 
between 410 and 500~keV agree well with those found for the 511~keV line by
\citet{Knoedlseder05}. Our bulge parameters are similar to, albeit
slightly smaller than, those obtained by \citet{Kinzer01} for the
positron annihilation radiation; our Ps continuum map is bulge
dominated as are the OSSE maps obtained by e.g.\
\citet{Milne01b,Milne02}.
%The all-sky distribution of extended emission in three
%energy bands between 410 and 500~keV. The extent and shape of the
%emission agree well with that found for the 511~keV line by
%\citet{Koedlseder05}.
The emission appears to be diffuse, we do not yet find evidence for
significant emission in the Ps analysis intervals from point sources
in the GC region. Modelling the extended emission with a Gaussian
bulge of $8^\circ$ FWHM located at the GC, and a disk component
tracing the Galactic CO distribution, we derive spectra and fluxes
that are consistent with previous findings obtained using SPI
\citep{Churazov05, Knoedlseder05}, OSSE
\citep{Purcell96, Kinzer99, Kinzer01}, and SMM \citep{Harris90}
%, and TGRS \citep{Harris98} 
observations. 
%{\bf ??? mention earlier ???}
%For example, we derive a Ps fraction of $f_{Ps} = 0.99 \pm 0.10$. 
%SPI therefore not
%only allows us to measure the total flux from our Galaxy, including
%faint and extented emission, but at the same time allows us to
%localize this flux better than any previous instrument. 

Emission between 410--500~keV from around the GC region is dominated
by the Ps continuum. Hence the sky distribution of Ps continuum
emission and 511~keV line emission
\citep{Knoedlseder05} are consistent, which is not surprising
considering that most positrons annihilate through formation of
Ps. This agreement strengthens our conclusions regarding the
origin of Galactic positrons based on observations of the 511~keV line
\citep{Knoedlseder05}. In particular, our results suggest that an old
stellar population, such as Type~Ia supernovae, classical novae, or
low-mass X-ray binaries, is the main source of Galactic
positrons. Light dark matter annihilation and/or decay is another
intriguing but speculative positron source.


\begin{thebibliography}{}

\bibitem[Bouchet et al.(2005)]{Bouchet05}
Bouchet, L., et al., 2005, ApJ, 635, 1103

\bibitem[Boehm et~al.(2004)]{Boehm04} 
%Boehm, C., Hooper, D., Silk, J., et~al., 
Boehm, C., et~al., 
2004, Phys.\ Rev.\ Lett.\ 92, 101301

\bibitem[Brown \& Leventhal(1987)]{Brown_Leventhal87} Brown, B.L., \&
Leventhal, M., 1987, ApJ, 319, 637

\bibitem[Cass{\'e} et~al.(2004)]{Casse04} 
%Cass{\'e}, M., Cordier, B.,Paul, J., et~al., 
Cass{\'e}, M., et~al., 2004, ApJ, 602, L17

\bibitem[Chan \& Lingenfelter(1993)]{Chan_Lingenfelter93} Chan,
K.-W., and Lingenfelter, R.E., 1993, ApJ, 405, 614

\bibitem[Chen et al.(1997)]{Chen97}
Chen, L.X., et al., 1997, ApJ, 481, L43

\bibitem[Churazov et~al.(2005)]{Churazov05} Churazov, E., et~al., 2005,
MNRAS, 357, 1377

\bibitem[Courvoisier et~al.(2003)]{Courvoisier03} Courvoisier,
T.J.-L., et al., 2003, A\&A, 411, L53

\bibitem[Dame et~al.(1987)]{Dame87} Dame, T.M., et al., 1987, ApJ,
322, 706

\bibitem[Dermer \& Murphy(2001)]{Dermer_Murphy01}
Dermer, C.D., \& Murphy, R.J., 2001, in {\sl Proc.\ of the
4$^{th}$ INTGERAL Workshop} (ESA SP-459), 115

\bibitem[Dickey \& Lockman(1990)]{Dickey_Lockman90}
Dickey, J.M., \& Lockman, F.J., 1990, ARA\&A, 28, 215

\bibitem[Dorman \& Arnaud(2001)]{Dorman_Arnaud01}
Dorman, B., \& Arnaud, K.A., 2001, in {\sl Astronomical Data Analysis
Software and Systems X}, ASP Conference Proceedings 238, 415
%in Astronomical Data Analysis Software and Systems X, ASP 
%Conference Proceedings, Vol. 238. Edited by F. R. Harnden, Jr., Francis 
%A. Primini, and Harry E. Payne. San Francisco: Astronomical Society of 
%the Pacific, p.415"

\bibitem[Forrest et al.(1980)]{Forrest80} Forrest, D.J., et al., 1980, Solar
Phys., 65, 15

\bibitem[Guessoum, Jean, \& Gillard(2005)]{Guessoum2005} Guessoum, N.,
Jean, P., \& Gillard, W., 2005, A\&A, 436, 171
 
\bibitem[Harris(1997)]{Harris97} Harris, M.J., 1997, in {\sl Proc.\ of the
Fourth Compton Symposium} (AIP 410), 418

\bibitem[Harris et al.(1990)]{Harris90} Harris, M.J., et al., 1990, ApJ, 362,
135

%\bibitem[Harris et al.(1998)]{Harris98}
%Harris, M.J., et al., 1998, ApJ, 502, 55

\bibitem[Jean et al.(2003a)]{Jean03a} Jean, P., Vedrenne, G.,
Roques, J.-P., et al.\ 2003a, A\&A, 407, L55

\bibitem[Jean et al.(2003b)]{Jean03b} Jean, P., Vedrenne, G.,
Roques, J.-P., et al.\ 2003b, A\&A, 411, L107

\bibitem[Jean et al.(2006)]{Jean06} Jean, P., et al., 2006, A\&A, 445, 579

\bibitem[Johnson et al.(1993)]{Johnson93} Johnson, W.N., 1993, ApJS,
86, 693

\bibitem[Kinzer et al.(1999)]{Kinzer99} Kinzer, R.L., Purcell, W.R.,
\& Kurfess, J.D., 1999, ApJ, 515, 215

\bibitem[Kinzer et al.(2001)]{Kinzer01} Kinzer, R.L., et al., 2001,
ApJ, 559, 282

\bibitem[Kn\"odlseder et~al.(2003)]{Knoedlseder03} Kn\"odlseder,
J., et~al., 2003, A\&A, 411, L457

\bibitem[Kn\"odlseder et~al.(2005)]{Knoedlseder05} Kn\"odlseder,
J., et~al., 2005, A\&A, 441, 513

\bibitem[Lebrun et~al.(2004)]{Lebrun04}
Lebrun, F., et al., 2004, Nature, 428, 293

\bibitem[Lonjou et al.(2004)]{Lonjou04}
Lonjou, V., et al., 2004, in {\sl Proc.\ of 5$^{th}$ INTEGRAL Workshop},
ESA-SP 552, 129

\bibitem[Milne et al.(2000)]{Milne00} 
%Milne, P.A., Kurfess, J.D., Kinzer, R.L., et~al., 
Milne, P.A., et~al., 2000, in {\sl Proc.\ of The Fifth Compton Symposium}
(AIP 510), 21

\bibitem[Milne et al.(2001a)]{Milne01a} 
%Milne, P.A., Kurfess, J.D., Kinzer, R.L., et~al., 
Milne, P.A., et~al., 2001a, in {\sl Proc.\ of the 4th INTEGRAL
Workshop 'Exploring the Gamma-Ray Universe'}, ESA SP-459, 145

\bibitem[Milne et al.(2001b)]{Milne01b} 
%Milne, P.A., Kurfess, J.D., Kinzer, R.L., et~al., 
Milne, P.A., et~al., 2001b, in {\sl Proc.\ of Gamma 2001} (AIP 587), 11

\bibitem[Milne et al.(2002)]{Milne02} 
Milne, P.A., et~al., 2002, New Astron. Rev., 46, 553

\bibitem[Ore \& Powell(1949)]{Ore_Powell49}
Ore, A., \& Powell, J.L., 1949, Phys.\ Rev., 75, 1696

\bibitem[Parizot et al.(2005)]{Parizot05}
Parizot, E., et al., 2005, A\&A, 432, 889

\bibitem[Purcell et al.(1996)]{Purcell96} 
Purcell, W.R., et~al., 1996, A\&AS, 120, 389

\bibitem[Purcell et al.(1997)]{Purcell97} 
%Purcell, W.R., Cheng, L.-X., Dixon, D.D., et~al., 
Purcell, W.R., et~al., 1997, ApJ, 491, 725

%\bibitem[Roques et al.(2003)]{roques_spi_perf} Roques, J.-P., Schanne, 
%S., von Kienlin, A., et al. 2003, this issue

\bibitem[Shrader et al.(2004)]{Shrader04}
Shrader, C.R., et al., 2004, in
{\sl Proc.\ of 5th INTEGRAL Workshop}, ESA-SP 552, 901

\bibitem[Sizun et al.(2004)]{Sizun04}
Sizun, P., et al., 2004, in
{\sl Proc.\ of 5th INTEGRAL Workshop}, ESA-SP 552, 815

\bibitem[Skinner \& Connell(2003)]{Skinner_Connell_03} Skinner, G.K., \&
Connell, P., 2003, A\&A, 411, L123

\bibitem[Strong et al.(2004)]{Strong04} Strong, A.W., et al., 2004, in
{\sl Proc.\ of 5th INTEGRAL Workshop}, ESA-SP 552, 507

\bibitem[Strong et al.(2005)]{Strong05} Strong, A.W., et al., 2005,
A\&A, 444, 495

\bibitem[Sturner et al.(2003)]{Sturner03}
Sturner, S.J., et al., 2003, A\&A, 411, L81

\bibitem[Teegarden et al.(2005)]{Teegarden05}
Teegarden, B.J., et al., 2005, ApJ, 621, 296

\bibitem[Tueller(1992)]{Tueller92}
Tueller, J., 1992, in {\sl Compton Gamma-Ray Observatory} (AIP 280),
97

\bibitem[Vedrenne et al.(2003)]{Vedrenne03} Vedrenne, G., et~al.,
2003, A\&A, 411, L63

\bibitem[Weidenspointner et al.(2003)]{Weidenspointner03}
Weidenspointner, G., et~al., 2003, A\&A, 411, L113

\bibitem[Weidenspointner et al.(2004)]{Weidenspointner04}
Weidenspointner, G., et~al., 2004, in {\sl Proc.\ of 5th INTEGRAL
Workshop}, ESA-SP 552, 133


\end{thebibliography}
\end{document}